\documentclass[twocolumn,runningheads,natbib]{svjour2}
\smartqed  
\usepackage{graphicx}

\journalname{Astrophysics and Space Science}
\newcommand{\be}{\begin{equation}}
\newcommand{\ee}{\end{equation}}

\def\X{{\mathrm{x}}}
\def\Y{{\mathrm{y}}}

\def\x{{\mathrm{x}}}

\def\B{\mathrm{B}}

\def\n{{\rm n}}
\def\p{{\rm p}}
\def\e{{\rm e}}

\def\c{{\rm p}}

\begin{document}

\title{Modelling the dynamics of superfluid neutron stars}

\author{N. Andersson}

\institute{School of Mathematics \\
University of Southampton \\ 
UK\\
              Tel.: +442380594551\\
              Fax: +442380595147\\
              \email{na@maths.soton.ac.uk}    
}

\date{Received: date / Accepted: date}

\maketitle

\begin{abstract}
In this brief summary I describe our recent work on superfluid neutron 
star dynamics. I review results on shear viscosity, hyperon bulk viscosity, 
vortex mediated mutual friction and the modelling of multifluid systems 
in general. For each problem I provide a set of questions that need to be addressed by future work.  
\end{abstract}

\section{Introduction}
\label{intro}

As the temperature of a system decreases towards absolute zero, matter 
either freezes to a solid or becomes superfluid. The first case 
is dominated by a structured lattice. In the second case, the 
system acts as a macroscopic quantum system. Regardless of the outcome, the 
rules that apply in an extremely cold system may be very different from those 
that describe the system at higher temperatures. This is, of course, a 
well-known fact and the different possible 
phases of matter have been studied in great 
detail. In particular, increasingly sophisticated laboratory experiments 
continue to provide insights into the details of superfluid/superconducting 
systems, ranging from the standard Bose-Einstein condensates to superfluid 
Helium \citep{donnelly} 
and atomic systems exhibiting fermion Cooper pairing. This area of 
research is highly relevant for those that are fascinated by neutron star
physics. In fact, mature neutron stars may provide the ultimate 
testing ground for theoretical physics. They are expected to contain
an elastic nuclear crust permeated by superfluid neutrons. An outer core
where superfluid neutrons coexist with superconducting protons transitions to 
the deeper core which may contain exotica like superfluid hyperons and 
deconfined quarks in a colour superconducting state. As if this was 
not enough, the presence of a ``solid'' core remains a possibility (even though
the relevant lattice parameters are 
unknown).    

Trying to understand the details of the 
various neutron star phases and their possible impact on 
astronomical observations is an exciting challenge. First of all, it requires 
a working knowledge of much of modern theoretical physics. Secondly, one 
will need some understanding of the observations. This includes decades of 
data for radio pulsars, in particular associated with the glitch events that 
provide the strongest  evidence of the presence of (at least partially) 
decoupled superfluid components. X-ray observations of accreting neutron stars
in binary systems, and isolated neutron stars closer to the Earth, are 
common and the data for the existence of magnetars is now very convincing. 
In fact, the recent observations of likely crustal oscillations following 
the giant flares in SGR1806-20 and SGR1900+14 [see the contributions by  Israel, 
 Watts and  Samuelsson] may be the harbingers of neutron star 
asteroseismology. Future gravitational-wave observations should add another 
observational window, although it will likely require an advanced generation 
of detectors sensitive at  high frequencies.

As the observations continue to improve we will have more 
precise quantitative opportunities
to test our theoretical models. To meet these tests the current models have to 
be improved. In fact, many of our ``favourite'' 
models are far too rudimentary to pass any closer scrutiny. 
This is not  surprising given that 
the relevant theory problems are difficult. Consider for example the 
possibility of superfluid vortex pinning in the neutron star crust.      
At the phenomenological level, the association between the
observed spin-up of the crust in a glitch and the transfer of angular momentum 
from an initially faster spinning (superfluid) component is natural. 
To invoke vortex unpinning as the key agent that motivates the 
event is also natural. Yet it is clear from the  effort
that has  gone into glitch modelling that it is  difficult to 
describe the process in a truly quantitative way.

The dynamics of a superfluid neutron star is associated with a number of similar
problems that need to be better understood. The main aim of the 
research described in this brief review is to arrive at 
such an understanding. Right now we are quite far away from this
goal. In some cases, e.g.  problems concerning basic shear viscosity, 
we may have a clear picture already, but in others, eg. how the
calculated shear viscosity coefficients are used in the case of 
a superfluid system,    
we are only beginning to be able to pose (what might be) the 
right questions. 

This brief review is a summary of my presentation at the London neutron star meeting. 
It is in no sense a complete review of the 
various problems. Rather it is a self-centered view of my thinking at the time of writing. 
The interested reader will find references to the relevant literature in the cited 
papers. 

Before I discuss our  recent work, let me
provide a list of problems that I find particularly interesting:

How do pulsar glitches really work? As already alluded to above, this 
remains a vexing issue. In my (somewhat pessimistic view) view, the present models are not 
truly quantitative. They likely contain the right elements, but the 
details of the underlying physics are not yet agreed upon. It is also worth
noting that most models predict how the glitch proceeds and how the system
relaxes back to quasi-equilibrium. Very few models provide a real 
mechanism to explain why the glitch happened in the first place. 

How do we understand neutron star free precession? The observational 
indications that some neutron stars are wobbling now seem relatively 
strong. The obvious question is why this behaviour is so rare. After all, 
free precession is the most natural mode of motion for any rotating body. 
The answer will provide insight into the damping of fluid motion in the star, 
and thus may constrain our models for internal dissipation. There could also 
be implications for vortex pinning or vortex-flux tube interactions in the 
superfluid/superconducting core [see Link's contribution]. 
It would also be nice to have a model 
for the excitation of the precession in the first place.  

Do the superfluid degrees of freedom affect the oscillation 
properties of the star? This is an important question that impacts on future 
attempts to probe neutron star physics via asteroseismology. To provide a 
``useful'' answer we need to study the problem using realistic models for
the superfluid pairing gaps (which, of course, are not agreed upon by 
theorists) within general relativity. The latter is key in order to make the 
results accurate enough that it is meaningful to compare to observations. 
In principle, I don't think the mode calculation would be very difficult.
At least not for non-rotating stars, and as long as one is prepared to accept that the
pairing gaps are likely to remain unknown up to perhaps a factor of a few.
However, if one  wants to 
account for the presence of the crust (which should be penetrated by 
superfluid neutrons) and the magnetic field, then the modelling becomes 
much more challenging. In the case of the crust, we have not (until very 
recently \citep{lars}) had any theoretical formulations that allow for the presence of a
superfluid component, while in the magnetic field case we do not have a good 
understanding of the internal field structure.     

Neutron stars may radiate detectable gravitational waves through a number of scenarios. One possibility is that the inertial r-modes are driven unstable
as they radiate gravitationally. The basic mechanism behind this instability is
 well understood, and a large number of damping mechanisms that 
counteract the instability have been considered. Thus it has become clear that the key
deciding factors relate to superfluidity. The presence of hyperons in the deep
stellar core may prevent the instability from happening, but if the 
hyperons are superfluid then the chemical reactions that lead to 
bulk viscosity are suppressed and  the effect on the instability may not be 
considerable. Another important mechanism is the so-called mutual friction, 
which in this context relates to the damping due to the scattering of electrons
off of magnetic fields associated with the superfluid neutron vortices. 
So far all studies of mode-damping due to this effect have been based on 
straight vortices (as in a rigidly rotating star). But
this may not be the appropriate model! If 
there is a significant oscillation in the star, which has some vorticity associated with it, then it seems likely 
that the vortices will get tangled up. The system would then be in a ``turbulent'' state, and from the analogous problem in superfluid Helium \citep{donnelly} 
we know that the 
mutual friction force is then rather different.  

Finally, I would like to emphasise the multi-fluid aspects of these
problems. In most available studies, concerning for example viscosity, 
it is implicitly assumed that we can apply the standard single fluid equations
of motion. Yet a multi-fluid system has extra degrees of freedom (the second 
sound in Helium and the analogous ``superfluid modes'' 
in an oscillating neutron star). These are unlikely to be ``passive''. In fact, 
we know that the equations that are used to model simple superfluid neutron 
stars admit a so-called two-stream instability \citep{2stream}. One can speculate that this
instability becomes relevant when the two components of the star rotate at 
different rates, as in the case of a spinning down neutron star with a 
pinned superfluid component. The instability mechanism is very simple and 
familiar from other physical systems, and it would be  interesting 
to know whether it can operate in a neutron star as well.

\section{Shear viscosity}
\label{sec:shear}

We have recently investigated the effects of superfluidity on
the shear viscosity in a neutron star core \citep{nucl}. We were motivated to do this 
by a puzzling result from the literature. The available results
suggest that the shear viscosity is \underline{stronger} 
in a superfluid neutron star
than it is when the star forms a normal-fluid system \citep{cutler}. This 
result contradicts our experience from other 
superfluid systems like Helium, and since the shear viscosity 
affects mode damping, free precession and relaxation 
after spin-up events in a crucial way, it is important that we 
understand it. 

To address the problem we combined existing theoretical results 
for the viscosity coefficients
with data for the various superfluid energy gaps into a ``complete'' 
description.
This lead to a simple model for the electron viscosity
which is relevant both when the protons form a normal fluid and
when they become superconducting. This turns out to be the key distinguishing 
factor. 

Below the neutron superfluid transition temperature the dominant 
contribution to the shear viscosity comes from the scattering of 
relativistic electrons. To model the shear viscosity we need to consider two
 contributions. 
Above the transition temperature at which the protons become superconducting, 
electron-proton scattering leads to a viscosity coefficient
\be
\eta_{\e\p}  \approx 1.8 \times 10^{18} \left( { x_\p \over 0.01 } \right)^{13/6} 
\rho_{15}^{13/6}  T_8^{-2} \mbox{ g/cm s}
\label{eta_est}
\ee
where $x_\p$ is the proton (electron) fraction, $T_8 = T/10^8 \mathrm{ K}$
and $\rho_{15} = \rho/ 10^{15} \mathrm{ g/cm}^3$.
Meanwhile, when the protons are superconducting the dominant effect is 
due to electrons scattering off of each other. Then we have
\be
\eta_{\e\e} \approx 4.4 \times 10^{19} \left( { x_\p \over 0.01 } \right)^{3/2} 
\rho_{15}^{3/2}  T_8^{-2} \mbox{ g/cm s}
\label{eta_ee}\ee
The protons play the key role since individual 
scattering processes add like ``parallel resistors''. That is, we have
\be
\tau = \left[ { 1 \over \tau_{\e\e} } + {  1 \over \tau_{\e\p} }\right]^{-1} \qquad \mbox{ where } \qquad \tau_{\e\e} >> \tau_{\e\p}
\label{tautot}\ee
and it is clear that the most important contribution
comes from the most frequent scattering process.

When the protons are 
superconducting the electron-proton scattering will be 
suppressed, essentially because there will be fewer states available for the
protons to scatter into. In order to allow for the transition to 
proton superconductivity, we can introduce a 
suppression factor ${\cal R}_\p$ such that
\be
\tau_{\e\p} \longrightarrow {\tau_{\e\p} \over {\cal R}_\p} 
\ee
Far below the critical transition temperature at which the protons become 
superconducting we should have ${\cal R}_\p\to 0$, and we see from  (\ref{tautot}) that the 
electron-electron scattering then dominates the shear viscosity.

\begin{figure}
\centering
  \includegraphics[width=0.8\columnwidth,clip]{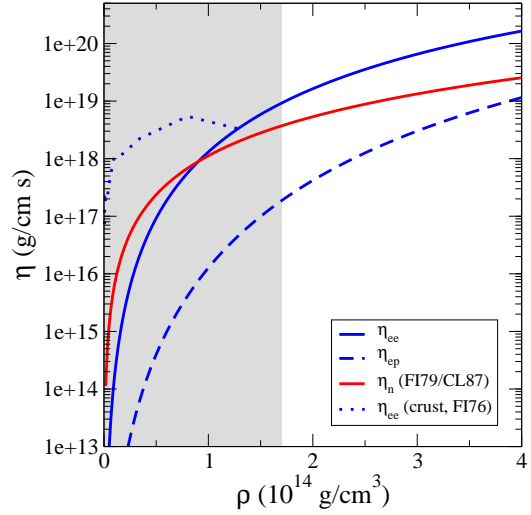}
\caption{Various shear viscosity estimates for a neutron star at
temperature $10^8$~K. 
The  solid blue line shows 
our estimate $\eta_{ee}$ while the  dashed blue curve corresponds to 
$\eta_{ep}$. 
These estimates are compared to i) the total shear viscosity in a normal
fluid neutron star, which is dominated by neutron-neutron scattering (FI79/CL87, solid red line) taken 
from  \citet{cutler}, and 
ii) results obtained by Flowers and Itoh  for the electron-electron shear viscosity in the crust
region (crust, FI76,  dotted blue curve). The crust  is indicated by the grey region. [See \citet{nucl} for the relevant references.]  } 
\label{ee-fig}       
\end{figure} 

Our  results, which are 
illustrated in Figure~\ref{ee-fig}, explain in a clear way
why proton superconductivity leads to a 
significant strengthening of the shear viscosity.
 It should be noted that the superfluidity of the 
neutrons is not the factor which leads to electron-electron
scattering becoming the main  shear viscosity agent. 
Rather, it is the fact that the onset of superconductivity
suppresses the electron-proton scattering. 
As can be seen from Figure~\ref{ee-fig}, the electron-electron shear
viscosity is not too different from the result for neutrons scattering off
of each other. This means that, in the temperature range where 
shear viscosity dominates, the  damping 
of neutron star oscillations will be quite similar (modulo multi-fluid effects) in the extreme cases 
when i) the neutrons and protons are both normal, and ii) when the 
neutrons and protons are superfluid/superconducting, respectively.
The contrast with the case when the neutrons are superfluid and the 
protons normal (and viscosity is dominated by $\eta_{\e\p}$) is clear from Figure~\ref{ee-fig}. This is an 
 interesting observation because it shows that proton 
superconductivity (or rather absence thereof) 
could have a significant effect on the dynamics of a neutron star core. 

{\em Future work needs to i) implement the different degrees of freedom of a multifluid system, and ii) determine the required 
suppression factors (like ${\cal R}_\p$).}

\section{Hyperon viscosity}

The presence of hyperons is expected to
affect a neutron star in several important ways. 
First of all, the $\Sigma^-$ carries negative charge which means
that the lepton fractions  drop significantly following its appearance. 
In fact, in some models there are virtually no electrons present 
in the core of the star. This may have  implications for the 
shear viscosity results that we discussed earlier.
Secondly, the hyperons can act as an extremely efficient refrigerant.
This is mainly because the hyperons may undergo direct URCA reactions 
essentially as soon as they appear, in clear
contrast to the protons which must exceed a threshold value
of $x_\p\sim 0.1$.   
Studies of the effect of this enhanced cooling 
have shown that a neutron star with a sizeable hyperon core would cool extremely fast. 
In fact, it would cool so fast that 
it would be much colder than observational data  suggests. This could be taken as evidence against 
an exotic core, but more likely it suggests that 
the hyperons are (at least partly) superfluid.
This is a natural explanation since
superfluidity leads to a reduction of the relevant reaction rates.

\begin{figure}
\centering
  \includegraphics[width=0.8\columnwidth,clip]{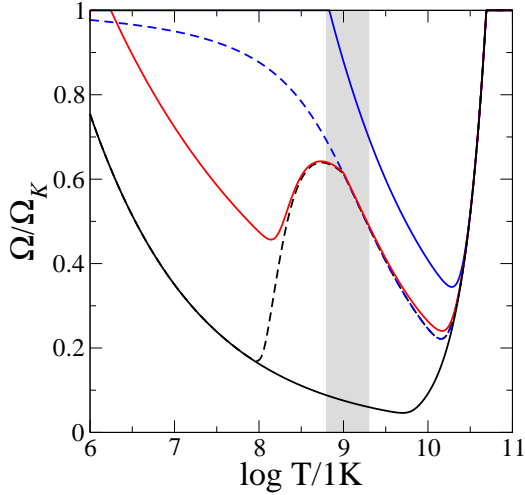}
\caption{Various estimates of the r-mode instability for a simple 
neutron star model with a 
hyperon core. The figure shows the critical rotation rate $\Omega/\Omega_K$ as a function of the core temperature $T$.  
The solid blue curve represents the
case of normal fluid hyperons. The r-modes are unstable above the curve, and 
even though there appears to be a sizeable instability window the hyperon 
viscosity provides a strong suppression (compare to the solid black curve which represents the case when there are no hyperons in the star). 
Basically, the remaining instability 
window is 
at such high temperatures that the star would cool through the instability 
regime in a minute or so. Far too fast to allow the modes to grow to a large 
amplitude. The dashed blue curve accounts for rotational effects, and shows that
the modes are unstable in the fastest spinning stars also at much lower 
temperatures. The solid  red and dashed black curves assume that the 
hyperon viscosity is suppressed below a given critical temperature 
(here taken to be  low, at  $10^9$~K), and correspond to the case when 
there is a crust (and a viscous Ekman layer is in operation) and
the star is entirely fluid, respectively.
}
\label{hypwin}       
\end{figure}

Analogous non-leptonic reactions have recently been invoked as a 
mechanism for producing a very strong bulk viscosity that may 
suppress the gravitational-wave driven instability  of the r-modes \citep{owen1}
[see also the contribution by Chatterjee \& Bandyopadhyay].
This mechanism is particularly important because, in contrast to the
standard bulk viscosity associated with $\beta$-reactions it is relevant at low
temperatures (the coefficient scales as $T^{-2}$ rather than $T^6$). 
For a while it was argued that the presence of hyperons would 
make the r-mode instability completely irrelevant. This is now understood 
not to be the case \citep{owen2}. Basically, the hyperons 
 must be superfluid in order not to lead to conflicts with cooling data.
Then the nuclear reactions that lead to the bulk viscosity 
must also be suppressed, and the effect on the 
r-mode instability may not be so great after all, see Figure~\ref{hypwin}.
In fact, the
most recent estimates suggest that the 
r-modes may be unstable in accreting neutron stars in LMXBs,
possibly regulating the spin of these stars. Since the associated 
gravitational-wave signal may be detectable by advanced detectors, it
is important that we understand this problem better. 

{\em Future work  needs to i) improve our understanding of the hyperon 
pairing gaps and the associated suppression factors for bulk viscosity, 
and ii) develop a multifluid description of bulk viscosity that can be used below the relevant superfluid transition temperature.}

\section{Modelling multifluids}

The equations that govern a general conservative multifluid system
can be derived from a constrained variational principle \citep{prix,monster}.  
The fundamental variables in this framework are the 
number densities $n_\X$, where $\x$ is a ``constituent'' index
that identifies the different particle species, and the associated 
transport velocities $v_\X^i$.
To complete the model one must provide an energy functional which represents
the equation of state. In the isotropic (meaning that there are no prefered directions)
two-fluid problem 
this functional takes the form $E(n_\n,n_\p,w^2)$
where $w_i^{\Y\X}\equiv v^\Y_i - v^\X_i $ ( $\X\neq\Y$ 
are constituent indices). This immediately leads to 
\be
dE = \sum_{\X=\n,\p} \mu_\X dn_\X + \alpha dw^2 
\ee
where $w^2 = w_i^{\Y\X} w^i_{\Y\X}$ and
\be
\tilde{\mu}_\X = { \mu_\X \over m_\B} = 
{1 \over m_\B} {\partial E \over \partial n_\X}\quad \mbox{ and } \quad
\alpha = {\partial E \over \partial w^2}  
\ee
(with $m_\B$ the baryon mass)
defines the chemical potential per unit mass $\tilde{\mu}_\X$
and the entrainment parameter $\alpha$. 

The
equations of motion for the  system can be written
\be
\partial_t n_\x + \nabla_i (n_\x v_\x^i) = 0 
\ee
and
\begin{eqnarray}
 n_\x \left(\partial_t + \pounds_{v_\x}\right) p_i^\x + n_\x \nabla_i \left(\mu_\x - 
        \frac{1}{2} m_\B v_\x^2\right)   
= 0
\label{euler}
\end{eqnarray}
where $\pounds_{v_\x}$ represents the Lie derivative along $v_\X^i$ (see \citet{living}  
for an explanation why the Lie derivative is natural in this context). 
In (\ref{euler}) the momentum (per particle) is given by 
$p_i^\x = m_\B ( v_i^\x + {\varepsilon_\x} w_i^{\Y\X} ) $, where $\varepsilon_\x = 2 \alpha/m_\B n_\x$. From this we can understand the nature of the 
entrainment effect. It is such that the velocity and momentum of each 
constituent in the multifluid system are no longer parallel. An alternative, 
perhaps more intuitive, description would represent this mechanism in 
terms of an   effective mass $m_\X^*$ \citep{mutual}. In the case of the outer core of a neutron star, where superfluid neutrons coexist with superconducting protons
the entrainment arises because of the strong interaction. Each proton/neutron 
is endowed with a cloud of particles of the opposite species, and when it flows it drags some of the other fluid along with it thus affecting
its momentum. 

The entrainment is known to play an important role in neutron star dynamics. 
In particular, it has been shown to affect the spectrum of stellar 
pulsation \citep{prix2,gusakov}. Moreover, it is one of the key ingredients
in the prescription for the  vortex mediated mutual friction. 

{\em Future work needs to
i) provide the entrainment parameters at finite temperatures (as in \citet{haensel})
(essentially doing equation of state calculations ``out of equilibrium''), 
and ii) develop fluid models that allow for the presence of both superfluid
condensates and the quasiparticle excitations that will be present 
when the star is no longer at $T=0$.}  

\section{Mutual friction}

A superfluid mimics bulk rotation by forming an array of vortices. The dynamics
of these vortices, and their interaction with the different particle species
may significantly affect the evolution of the system. In a neutron star
the presence of vortices leads to ``mutual friction'' 
between interpenetrating superfluids (neutrons and protons).
This couples the two fluids on a relatively short timescale, which has
implications for both glitch recovery estimates and the damping of 
pulsation modes. In fact, an estimate by Lindblom and Mendell indicates that 
the gravitational-wave driven instability of the fundamental f-modes will 
be prohibited by the mutual friction \citep{lm1}. The same effect is also 
relevant, albeit not quite in such a dramatic fashion, for the r-modes \citep{lm2}.

\begin{figure}
\centering
  \includegraphics[width=0.6\columnwidth,clip]{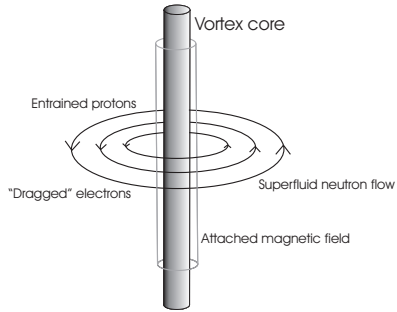}
\caption{A schematic illustration of the flow around a superfluid 
neutron vortex. Because of the entrainment effect, the vortex induces flow also in the protons. This leads to a magnetic field forming on the vortex, and the 
scattering of electrons off of this magnetic field leads to a
dissipative mechanism known as mutual friction. }
\label{entrain}       
\end{figure} 

In a recent study, we have revisited the problem of the mutual friction
force for neutron stars \citep{mutual}. It is known that the entrainment 
plays a key role in determining the strength of the 
mutual friction, and to make progress and study 
various astrophysical scenarios we needed to develop a 
description of this mechanism within our multifluid framework. 
Our results  provide useful checks on the classic work by \citet{alpar1} (who were the first to discuss the problem) and 
 \citet{mendell} (whose mutual friction coefficients were used in the studies of 
mode-instabilities in superfluid neutron stars).

Briefly, the mutual friction that is expected to be the most important
in the outer core of a mature neutron star originates as follows. 
The superfluid neutrons form vortices, which represent the quantisation
of the momentum circulation. That is, we have
\be
\kappa^i = { h \over 2 m_\n} \hat{\kappa}^i = \epsilon^{ijk} \nabla_j p_k^\n
\ee
where $\hat{\kappa}^i$ is a normalised tangent vector to the vortex.
Here it is important to understand that it is the circulation of momentum that is 
quantised, not that of the velocity. At first sight this distinction may seem to suggest that our model differs from the standard picture. However, this would be wrong. One must remember that in the orthodox description for superfluid Helium, the 
so-called superfluid ``velocity'' is in fact a rescaled momentum \citep{prix}. Hence, the two pictures are consistent, the only difference being that we  make the 
true dynamical role of the variables clearer. Now, the flow of neutrons associated with the vortex
induces flow in part of the protons because of the entrainment. This leads to the generation of a magnetic field, see Figure~\ref{entrain},
the strength of which is estimated as 
\be
B \approx 2\times10^{14} \mathrm{G} \ \left( { x_\p \over 0.05} \right) \rho_{14}  \left| { m_p -m_p^* \over m_p^*} \right| 
\ee
where  $\rho_{14} = \rho/ 10^{14} \mathrm{ g/cm}^3$.
Finally, electrons 
 scatter dissipatively off of this magnetic field, leading to
a coupling of the two fluids in the system (the superfluid neutrons and a
conglomerate of all the charged particles).

To arrive at the standard expression for the mutual friction force, we 
balance the Magnus ``force'', which acts on a vortex as the superfluid 
flows past it,  and a resistive force due to 
electrons scattering off of the vortex. It should be noted that the 
Magnus effect is already present in the smooth averaged equations of motion (\ref{euler}), which means that in practice we  only add resistivity to the 
right-hand side of the equations.
Anyway, the force-balance leads to the mutual friction force acting on the neutrons being
\be
f_\mathrm{mf}^i = {\cal B} \rho_n n_v \underbrace{\epsilon^{ijk} \hat{\kappa}_j \epsilon_{klm} \kappa^l w^m_{\n\c}}_{\mathrm{projects  } \perp \kappa^i } + 
{ {\cal B}^\prime} \rho_\n n_v \underbrace{\epsilon^{ijk} \kappa_j w_k^{\n\c}}_{\mathrm{acts }\perp  w_k^{\n\c}}  
\label{mf_final}\ee
This force has been used to model a number of astrophysical 
scenarios. In particular, 
the coupling of the two fluids following a glitch event, see the following 
section, and the damping of oscillations in a superfluid neutron star.  
It is, however, important to understand that this is only a first 
approximation to the real mutual friction interaction.  
There are presently two alternatives that should be taken seriously. 
The first, which has been advocated by Sedrakian and collaborators for 
over 20 years [see \citet{sedrakian} and references given therein], 
is based on the notion of vortex clusters. 
The idea is that each neutron vortex is associated with a large 
collection of proton fluxtubes. This  increases the scattering cross section for the electrons, and as a result the mutual friction force is
many orders of magnitude stronger than  given above. The 
upshot of this is (somewhat counter-intuitively) that the coupling between the fluids take place on a much 
longer timescale (months rather than seconds). The other alternative 
accounts for the curvature of the vortices, and the possibility 
that they will get tangled up in a state that in many ways 
is reminiscent of turbulence. The lack of a prefered direction of the vortex
array obviously affects the form of the mutual friction force. 
This is well-known from work on superfluid Helium \citep{donnelly}. Nevertheless, this
alternative form for the mutual friction has only recently been 
discussed for neutron stars. \citet{peralta} are leading the way by 
carrying out numerical simulations of a turbulent two-fluid system.
This work has a number of interesting implications, and 
it will be exciting to see where this will take us.

{\em Future work needs to investigate the alternative manifestations 
for the mutual friction in detail. In my opinion the most pressing 
issue concerns the turbulence. Since it is known to be the key effect 
in experiments on
superfluid Helium one has every reason to expect that it will be 
relevant for neutron stars as well.}  
 
\section{Dynamical coupling}

One of the most important problems that can be addressed once 
we know the form of the mutual friction force regards the relaxation
timescale following a pulsar glitch. The idea is that in the 
initial state the vortices are pinned. This leads to a build-up of
a rotational lag between the two fluids as the charged component
(the crust and the core protons/electrons) spins down. At some critical relative rotation rate, the pinning force will no longer be able to hold the vortices in place. Global unpinning then leads to evolution described by the equations 
that we have derived (provided that the vortices remain straight during the evolution, a debatable assumption). All we need in order to be able to estimate the relaxation timescale following the glitch are the coefficients in (\ref{mf_final}). 
 As we have recently shown \citep{mutual}, 
the dimensionless coefficients are ${\cal B}^\prime = {\cal B}^2$ and
\begin{eqnarray}
{\cal B} &=& { { \cal R} \over \rho_\n \kappa} \approx \nonumber \\
&\approx&  
4\times10^{-4} \left( {\delta m_\p \over m_\p} \right)^2
\left( {m_\p \over m_\p^*} \right)^{1/2} \left( { x_\p \over 0.05} \right)^{7/6}  \rho_{14}^{1/6} 
\end{eqnarray}
where $\delta m_\p = m_\p - m_\p^*$. 
Working out the dynamical coupling timescale we easily find 
\be
\left. \begin{array}{ll} 
n_\n \partial_t p_i^\n + \ldots = f^\mathrm{mf}_i \\
n_\p\partial_t p_i^\c + \ldots = - f^\mathrm{mf}_i \end{array} \right\} \rightarrow  {  m_\p^* \over m_\p} 
\partial_t w_i^{\n\c} + \ldots \approx  -  { { \cal B} \kappa n_v \over x_\p}  w_i^{\n\c}
\ee
That is, the timescale on which the two fluids are (locally) coupled can be 
estimated as
\begin{eqnarray}
\tau_d &\approx&  {  m_\p^* \over m_\p}  { x_\p \over {\cal B} \kappa n_v} 
\approx  \nonumber \\ &\approx& 10 P (s)
\left( { m_\p^*\over \delta m_p} \right)^2  \left( { x_\p \over 0.05} \right)^{-1/6} \rho_{14}^{-1/6}
\end{eqnarray}
This suggests that the two core fluids are coupled on a timescale of about 
10 rotation periods (for typical parameters). This estimate should be 
compared to the classic result of \citet{alpar2} who suggest 
that the coupling timescale is $400-10^4$ periods. 
This relatively rapid coupling is usually taken as evidence that glitches
must originate in the crust superfluid, not in the core. Our calculations, 
with a coupling that 
is about 1 order of magnitude \underline{faster} than the old result, 
obviously support this conclusion. However, we need to understand why 
our estimate differs from that of Alpar and Sauls. The result may seem surprising since the parameters in the mutual friction force are
 similar in the two studies.  The difference comes from the set-up of the coupling timescale problem. 
Our estimate followed immediately from the two-fluid equations, and  represents the timescale on which mutual friction 
couples the two components in the system. In contrast, Alpar and Sauls consider the 
equation of motion for the vortices. Setting the system into relative rotation, they work out the timescale for the vortices to relax to a new
equilibrium position. 
The key difference between the two estimates is that in the Alpar-Sauls
scenario the rotational lag between the fluids is maintained while the vortices relax. This means that their model does 
not conserve angular momentum during the relaxation, which explains why the two results are different.

\section{Assignments}

At the end of my presentation, I gave the audience 
a number of ``homework assignments''. These represent problems that I
feel need more attention, and which I hope to consider in
the near future. It seems appropriate to conclude this brief write-up 
by outlining these problems. 

As we are beginning to understand the basic dynamics of the two-fluid model 
for neutron stars, we should look ahead and develop a framework 
that would allow us to make our models more realistic. Most importantly, we
need to account for finite temperature effects and viscosity. 
This is a difficult problem area, but I feel that we cannot shirk 
away from it. Recently, Greg Comer and I have developed a 
flux-conservative formalism for multi-fluid 
systems, including both dissipation and entrainment \citep{monster}. Our formulation 
has much in common with work on the mechanics of mixtures and 
multiphase flows. Our first study of the problem is quite formal, and
should be relevant for any multifluid situation in Newtonian physics. 
Of course, we want to apply the framework to neutron stars. 
To do this, we have considered two model systems. The results make it painfully clear that  more realistic models will be significantly more complex than 
the systems we have considered so far. 
Our analysis shows that the simplest ``reasonable'' model for a neutron 
star core, starting with four fluids (neutrons, protons, electrons and entropy)
and reducing to three degrees of freedom (superfluid neutrons, entropy and everything else), requires no less than 19 dissipation coefficients. 

\begin{figure}[h]
\centering
  \includegraphics[width=\columnwidth,clip]{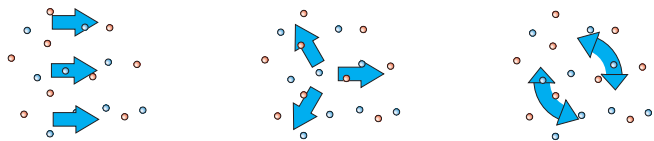}
\caption{An illustration of the different kinds of motion that lead to 
dissipation via particle scattering in a two-fluid problem (blue and red).
The ``red'' fluid is assumed to be at rest, and the ``blue'' fluid i) flows 
linearly, ii) contracts and expands, and iii) undergoes shearing motion.
}\label{coeff}       
\end{figure} 

It will be a challenge 
to understand the role and values of these different coefficients. Yet it 
may not be too bad a problem, because the meaning of many the of new coefficients is quite clear. Consider for example dissipation due to particle scattering. 
In a single fluid problem scattering leads to viscosity due to shearing motion in the flow. In a situation where a 
relative flow is possible, there are more degrees of freedom that lead to particle scattering. Consider the schematic illustration in Figure~\ref{coeff}, where the two fluids are represented by blue and red particles. 
There are  three different ways that relative flow can induce
scattering, and result in dissipation. Taking the ``red'' fluid 
to be at rest, the blue fluid can i) flow linearly, ii) contract and expand, or iii) undergo shearing motion. The viscosity coefficients associated with these different degrees of freedom are  related and should only differ by ``geometric factors''. They should certainly be calculable, and might actually be known from kinetic theory already.

As described in other contributions to these proceedings, it is 
likely that we now have the first observations of neutron star oscillations. 
This is fantastic news, but it provides us with a challenge that must be 
met in the near future. We need to develop accurate theoretical models
 to put contraints on eg. the crust physics via these observations. 
However, if we want to do this we must  
consider fully relativistic models. This is the only way to obtain results that have the precision required to make a comparison with observations meaningful. 
In fact, we might even have to account for the presence of superfluid neutrons in the crust, a very difficult problem. Moreover, the observed systems are 
all magnetars and one might expect the magnetic field to be important. 
It seems likely (at least to me) that the magnetic field will 
couple any motion in the crust to the core, thus altering the 
nature of the pulsation modes and making any analysis of ``pure crust
oscillations'' less relevant \citep{magnetic}. This would force us to 
study global oscillations of magnetised fully relativistic stars, certainly not a
simple problem. 

As I have already indicated, we need to worry about 
superfluid ``turbulence''. I would not be at all surprised if 
studies of this problem lead to results that change our understanding of
neutron star dynamics significantly. One advantage is that there 
has been  a lot of work on the analogous problem for superfluid Helium \citep{donnelly}, 
and we can  hope to benefit from these results.
Most importantly we need to understand whether a turbulent description 
is relevant for neutron stars. If it is, how does it manifest itself?
What is the effect on, for example, glitch relaxation and mutual friction 
damping of neutron star oscillations?
 
In addition to these problems, I can think of a number of  issues 
concerning multifluid aspects of exotic phases like hyperons and 
deconfined quarks. Certainly the neutron star community is bustling 
with exciting ideas, but  we have a lot to do before we truly understand the dynamics of superfluid neutron stars.

\begin{acknowledgements}
Much of the work described in this article was done in collaboration with 
Greg Comer. Without his contribution the results discussed here would not have been the same.
\end{acknowledgements}



\end{document}